\documentstyle[11pt]{article}

\begin{document}
\large

\def\lsim{\mathrel{\rlap{\lower3pt\hbox{\hskip0pt$\sim$}}
    \raise1pt\hbox{$<$}}}         
\def\gsim{\mathrel{\rlap{\lower4pt\hbox{\hskip1pt$\sim$}}
    \raise1pt\hbox{$>$}}}         
\def\dblint{\mathop{\rlap{\hbox{$\displaystyle\!\int\!\!\!\!\!\int$}}
    \hbox{$\bigcirc$}}}
\def\ut#1{$\underline{\smash{\vphantom{y}\hbox{#1}}}$}

\newcommand{\beq}{\begin{equation}}
\newcommand{\eeq}{\end{equation}}
\newcommand{\dem}{\Delta M_{\mbox{B-M}}}
\newcommand{\dega}{\Delta \Gamma_{\mbox{B-M}}}

\newcommand{\ind}[1]{_{\begin{small}\mbox{#1}\end{small}}}
\newcommand{\tind}[1]{^{\begin{small}\mbox{#1}\end{small}}}

\newcommand{\WA}{{\em WA}}
\newcommand{\SM}{Standard Model }
\newcommand{\QCD}{{\em QCD }}
\newcommand{\KM}{{\em KM }}
\newcommand{\hscale}{\mu\ind{hadr}}
\newcommand{\sG}{i\sigma G}

\newcommand{\MS}{\overline{\mbox{MS}}}
\newcommand{\pole}{\mbox{pole}}
\newcommand{\aver}[1]{\langle #1\rangle}

\newcommand{\appa}{\mbox{\ae}}
\newcommand{\CP}{{\em CP } }
\newcommand{\fy}{\varphi}
\newcommand{\hi}{\chi}
\newcommand{\al}{\alpha}
\newcommand{\as}{\alpha_s}
\newcommand{\gf}{\gamma_5}
\newcommand{\gm}{\gamma_\mu}
\newcommand{\gn}{\gamma_\nu}
\newcommand{\be}{\beta}
\newcommand{\ga}{\gamma}
\newcommand{\de}{\delta}
\renewcommand{\Im}{\mbox{Im}\,}
\renewcommand{\Re}{\mbox{Re}\,}
\newcommand{\GeV}{\,\mbox{GeV}}
\newcommand{\MeV}{\,\mbox{MeV}}
\newcommand{\matel}[3]{\langle #1|#2|#3\rangle}
\newcommand{\state}[1]{|#1\rangle}
\newcommand{\ra}{\rightarrow}
\newcommand{\ve}[1]{\vec{\bf #1}}

\newcommand{\rhs}{{\em rhs}}
\newcommand{\pp}{\langle \ve{p}^2 \rangle}

\newcommand{\BR}{\,\mbox{BR}}
\newcommand{\La}{\overline{\Lambda}}
\newcommand{\Lam}{\Lambda\ind{QCD}}

\newcommand{\np}{nonperturbative}
\newcommand{\re}[1]{ref.~\cite{#1}}

\vspace*{.7cm}
\begin{flushright}
\large{
CERN-TH.7091/93\\
UND-HEP-93-BIG\hspace*{0.1em}07}\\
November 1993
\end{flushright}
\vspace{1.2cm}
\begin{center} \LARGE
{Anathematizing the Guralnik-Manohar Bound for $\La$}
\end{center}
\vspace*{.4cm}
\begin{center} \Large
I.I.Bigi\\
{\normalsize\it TH Division, CERN, CH-1211 Geneva 23,
Switzerland \footnote{During the academic year 1993/94}}\\
and\\
{\normalsize\it Dept.of Physics,
University of Notre Dame du
Lac, Notre Dame, IN 46556, U.S.A.\footnote{Permanent address}}
\\{\normalsize \it e-mail address: VXCERN::IBIGI, BIGI@UNDHEP}
\vspace{.4cm}
\\
N.G.Uraltsev\\
{\normalsize\it TH Division, CERN, CH-1211 Geneva 23,
Switzerland}\\
and\\
{\normalsize \it St.Petersburg Nuclear Physics Institute,
Gatchina, St.Petersburg 188350, Russia $^2$}
\\{\normalsize \it e-mail
address:  VXCERN::URALTSEV, URALTSEV@LNPI.SPB.SU}
\end{center}
\thispagestyle{empty} \vspace{.4cm}

\centerline{\Large\bf Abstract}
\vspace{.4cm}
There is a recent claim by Guralnik and Manohar \cite{GM}
to have established a
rigorous lower bound on $\bar \Lambda$, the asymptotic
difference between the mass of a heavy flavour {\em hadron} and
that of the heavy flavour {\em quark}.
We point out the flaw in their
reasoning and discuss the underlying physical problem.
An explicit counterexample to the GM
bound is given; one can therefore not count on a refined
proof to re-establish this bound.

\newpage
\large
\addtocounter{footnote}{-2}

Very significant progress has been made in the last few years
in the theoretical
description of heavy flavour hadrons: in particular one has gained
more control over the non-perturbative effects
in static quantities as well as
in some decay processes. The progress in the latter
has come from an
extensive application of Effective Heavy Quark Theory (EHQT) \cite{EHQT}
to certain exclusive semileptonic modes on the one hand, and on the
other from a direct QCD treatment [3-7]
of inclusive decays through an expansion in powers of $1/m_Q$
with $m_Q$ being the heavy flavour quark mass.
In particular, the question of the leading
non-perturbative corrections to inclusive transition rates
has been addressed
in these QCD papers: it was shown there that corrections
of order $1/m_Q$ are
absent in many instances, for example in total widths;
they do however emerge in other cases,
namely in the shape of semileptonic spectra or even
in such integrated quantities like the average
invariant mass of the hadronic final state for semileptonic or radiative
decays \footnote{The problem of
$1/m_Q$
effects was briefly addressed already in ref.~\cite{CGG} with respect to
semileptonic decays only; however incorrect conclusions were stated about
the existence of these corrections.}.

There is one quantity of particular interest since it controls the
leading non-perturbative effects for some decay rates and at the same time
plays a fundamental
role in EHQT: it is usually referred to as $\La$ and defined as
the difference in the mass of a heavy flavour hadron $H_Q$
and of the corresponding heavy quark Q in the limit
$m_Q\ra \infty$:
$$ \matel{H_Q}{\bar Q \frac{1}{2}\,
(i\nabla_\mu\gn+i\nabla_\nu\gm) Q}{H_Q}=
(1-\La/M_{H_Q})\cdot \matel{H_Q}{T_{\mu\nu}}{H_Q}\eqno(1)$$
where $T_{\mu\nu}$ denotes the full energy-momentum tensor in QCD whereas
the operator on the {\em lhs} represents only
the part associated with the heavy
quark. The value of $\La$ is not known for sure even for the family of the
lightest pseudoscalar mesons; reasonable estimates center around the
value $\La_P \sim 300\div 500\MeV$.

It has been claimed in a recent paper by Guralnik and Manohar
(hereafter referred to as GM) that one can derive a
rigorous lower bound $\La \ge 237\MeV$ for the heavy flavour
pseudoscalar mesons. The aim of the present
note is to show that this conclusion is erroneous:
the actual proof given in ref.\cite{GM} is incorrect and its flaw
in all likelyhood cannot be
cured in any reasonable way. We will also
present an explicit counterexample to the GM bound.
Moreover we will give arguments that no
sensible {\em rigouros} bound can be obtained at all for $\La$; at the
same time some physically relevant statements can be made.

The idea underlying the derivation of \re{GM} is to use
existing QCD mass inequality methods \cite{W,N} for hadrons containing
heavy quarks. A
straightforward application of such inequalities yields according to
eq.(12) of GM the following bound:
$$M(\bar Q q) \ge
\frac{1}{2}[ m'(\bar Q i\ga_5 Q) + m'(\bar q i\ga_5 q)] \eqno(2)$$
(the prime indicates that the annihilation contribution
has been ignored, i.e. the
masses stand for hypothetical mesons with non-identical,
though mass-degenerate quarks and antiquarks);
the inequality holds irrespective of the quantum numbers
of the $\bar Q q$ state. This inequality {\em per se} does not lead to
any bound on $\La$ due to the dominant, $\sim \as^2\cdot m_Q$,
negative Coulomb energy of the $\bar QQ$ state on the
{\em rhs}.
Another inequality,
$$M(\bar Q q)- m_Q \ge   1/2\,m'(\bar q i\ga_5 q) \eqno(3)$$
was stated in eq.(11) of GM using
arguments based on the effective theory for
the infinitely heavy quark $Q$. This inequality
differs from the previous one
by the absence of the binding energy for the heavy quarkonium
state and represents the
main result of \re{GM}. We will show now that this bound is physically
irrelevant.

The problem can be formulated in short as follows: in the {\em effective}
theory of an {\em infinitely} heavy quark there exists no room
for a quantity $m_Q$;
therefore any results obtained in such a theory cannot be related to
the difference between the masses of hadrons and the mass of the heavy quark.
To be more specific: while the energy of a hadron state in the
effective theory can be measured, say, in lattice simulations, it has in fact
no direct connection to $\La$. Therefore the inequality stated in
eq.(10) of GM does not lead to an
inequality for $\La$ as claimed in eq.(11) of GM.

  To clarify this at first sight paradoxical comment
one needs to examine the subtleties of the derivation in
more details. More specifically we shall now discuss the question
of how the
inequality for the correlators of the heavy quark currents
$$|\aver{[\bar q \Gamma^+ Q](y)\,[\bar Q \Gamma q](x)}|^2 \le
C \cdot
\aver{[\bar q i\gamma_5 q](x)\,[\bar q \gamma_5 q](y)} \eqno(4)$$
that was given in eq.(10) of GM can be used -- and as
it turns out, actually cannot be used -- to arrive at eq.(3).
Eq.(4), when considered for
$\vec{x}=\vec{y}$ and a large Euclidean time difference $|x_4-y_4|\ra
\infty$ states that the energy of $\bar Q q$ states in the effective
theory is
not less than half the energy of the lowest lying $\bar q q$ state.
The question is how can this result be related to $\La$.

GM consider a lattice version of the theory originally
containing the `real' heavy quark
$Q$ with some finite mass $m_Q$. To obtain the $Q$ propagator
corresponding to the one in EHQT one then has to consider the limit
$$m_Q \gg 1/a \eqno(5)$$
with $a$ denoting the lattice spacing. Doing so one indeed
arrives at eq.(3).
However it is obvious that the mass $m_Q$ entering this equation
is the bare
mass, $m_Q^{bare}$, that was originally present in the lattice
theory. On the other hand
the mass that enters the definition of $\La$ is not the bare one but rather
the {\em pole} (or `on-shell') mass.

In real QCD the two masses, namely the pole and the bare
mass, differ by an infinite amount
$\sim \as/\pi \cdot m_Q \log{\Lambda\ind{uv}^2/m_b^2}$ where $\Lambda\ind{uv}$
denotes some ultraviolet cutoff. In lattice QCD
where there is a built-in maximal momentum scale of order $1/a$,
there is a finite difference between the two masses. In the limit under
consideration -- $m_Q \gg 1/a$ -- this difference is
simply given by the
classical Coulomb self-energy of a charged particle producing electrostatic
fields:
$$m_Q\tind{pole}-m_Q\tind{bare}
\simeq \frac{2}{3}\as/R \sim \as/a \eqno(6)$$
where $R$ is the `radius' of the charged particle, $R\sim a$; it is important
that the {\em rhs} of eq.(6) is necessarily positive. The bound for the
quantity of interest, $\La$ then reads as
$$\La_{\bar Q q} \ge  - c \,\as /a + 1/2\,m'(\bar q i\ga_5 q) \eqno(7)$$
with $c>0$, i.e. with an additional negative term on the {\em rhs}
relative to eq.(3).

How essential is this modification? The bound  in eq.(7) of GM
indeed is much stronger
than the inequality in eq.(2) obtained in the standard
way \cite{W,N}. However
to be able to incorporate the light
degrees of freedom of QCD one has to assume that $1/a\gg \Lambda\ind{QCD}$.
Then one realizes that the effective theory still gives
an inequality that is trivially
fulfilled: although the {\em rhs} of the equation does not scale like
$m_Q$, it still does not provide a useful lower bound: for the
first term
is parametrically larger than $\Lambda\ind{QCD}$ and negative!

At first sight it would seem
one can try to evade this problem by a natural modification
in the line of reasoning, namely by considering a smaller
lattice spacing in order to reach the continuum limit even
for the heavy quarks. Unfortunately that does not work
either, as revealed by closer scrutiny.  For as soon as the heavy quarks
become non-stationary through interactions with the gluon
fields that are non-leading in $1/m_Q$,
the average of the product of the two heavy quark propagators $W$ in
in eq.(7) of GM is no
longer bounded from above by a constant -- instead it grows exponentially
with time. These corrections are generated by extra terms
in the
heavy quark propagators that appear in eq.(3) of GM. This positive
exponent is a reflection of the (attractive) interaction between the
propagating heavy quarks that is mediated by the gluon field.
Then additional
negative terms appear in the {\em rhs} of the inequality for $\La$ and thus
the GM inequality gets invalidated. The origin
of the additional interaction
for propagating heavy quarks is very lucidly discussed in GM and we do not
need to address it here.

It was alluded in GM to a different treatment of the effective
theory that leads to the same inequality. Since that proof has not been
given in GM we cannot of course point out at which point
exactly it goes awry. Instead we will present a concrete
counterexample to the basic inequality. This will
demonstrate that this bound cannot be resurrected by a more
clever proof.

Let us consider muonium in QED, i.e. the boundstate of a muon
and a positron in a hydrogen-like system. This represents a
`Heavy Fermion' scenario with the added bonus that the
boundstate properties can be calculated explicitly; the muon
mass provides the high mass scale and $m_e$, the positron mass,
sets the scale for the `light degrees of freedom' in the EHQT
\footnote{Likewise one can also consider
real QCD with two heavy quarks, namely $b$ and $c$
with $m_b\gg m_c\gg \Lambda _{QCD}$.}. All the
general arguments employed in GM apply here directly and
therefore one would obtain the inequality of eq.(3).

The mass spectrum in such a theory is well known:
the boundstate mass equals the sum of the {\em pole}
masses of the constituents plus the Coulomb binding
energy $E_C=-\alpha ^2/2\cdot m_{red}$ where
$m_{red}$ denotes the reduced mass. Both sides of the
inequality can thus be evaluated leading to the
claim
$$m_\mu +m_e - \frac{\al^2}{2}\,m_e - m_\mu \ge
\frac{1}{2}\:(\,2m_e-\frac{\al^2}{2}\,\frac{m_e}{2}\,) \eqno(8)$$
which is obviously incorrect.

This explicit counterexample demonstrates that the
bound claimed in GM cannot be established
rigorously, independently of possible technical
modifications of the concrete
proof. We believe that this `no-go' statement is not accidental,
but reflects some deeper reasons.
Although our arguments are rather general and
cannot be considered as any kind of rigorous statement, we shall discuss them
now.

The main difficulty in making precise statements about $\La$ in terms of,
say, $\Lambda\ind{QCD}$ is that $\La$ is not well defined in
terms of the {\em bare}
parameters of QCD.  While the masses of hadrons constitute clearly
unambiguous quantities, the renormalized mass of the heavy quark
does not; for it suffers from
perturbative corrections having a {\em multiplicative} form:

$$m_Q\tind{pole}=m_Q+c_1\as(m_Q^2)\,m_Q+c_2\as^2(m_Q^2)\,m_Q+
\ldots\;\;\;;\eqno(9)$$
the generic term in the
series has then the form $m_Q/\log^n(m_Q/\Lam)$. This
series is asymptotic, and any term in it must be considered to be
parametrically larger than the quantity of interest, $\La\,$,
in the limit $m_Q\ra \infty$. From a formal point of view the
result for $\La$ thus depends
completely on the way how the infinite series is summed up.

The arguments given above
may seem to suggest that any constructive definition of
$m_Q\tind{pole}$ and therefore of $\La$ is impossible with an
accuracy \footnote{From now on we neglect factors
of $\log(m_Q)$.} of better than $\sim m_Q$.
The conventional argument would be that due to
colour confinement the asymptotic states corresponding to free quarks do
not exist; therefore whatever accurate method of treating QCD is
developed
in the future, the pole mass cannot be defined. We believe that such
agnosticism overstates the real problem. For one can consider that phase of
QCD where the gauge symmetry is spontaneously broken by some Higgs
fields at the relatively low scale $\gsim \Lam$ which is much smaller
than $m_Q$. The energy of the deconfined single heavy quark states do depend
of course on the details of the Higgs sector in such theories; however
the scale for the
variation in the mass is given by $\Lam$ and not by $m_Q$. The pole
mass of the
heavy quark defined in such a way must identically coincide
with the standard
definition in QCD to all orders of the perturbative expansion.

One sees therefore that a constructive definition of the $m_Q\tind{pole}$
can be given in a
straightforward way even beyond perturbation theory. However it
neccessarily suffers from an uncertainty of order
$\Lam$; that is just the relevant scale
for a discussion of $\La$. One of the possible
such definitions is of course to state that $m_Q\tind{pole}$ is
simply the mass of, say, the lightest pseudoscalar $\bar Q q$ state with
a massless
spectator. This definition satisfies all
necessary requirements and is of course completely constructive, however
the question about the size of $\La$ becomes a tautology.

One could think of somehow defining
the heavy quark pole mass just via the lattice realization of QCD
with heavy quarks as was implied in GM. Then the bound claimed
in that paper would be rigorous
for the pole mass defined in some specific way. Our analysis
of the GM proof given above shows, however,
that this definition can {\em not}
satisfy the necessary requirements: it yields a pole mass which is
{\em parametrically
smaller} -- by an amount $\sim 1/a \log(a)$ --
than any of the `correct' pole masses.
Therefore any finite inequality derived for a thus defined mass is
without physical
relevance.

We thus arrive at the conclusion that from a formal point of view
a description
of powerlike nonperturbative effects makes sense only
\cite{BUV,BU2,BU,BU3,BSUV2} for quantities that have no purely
perturbative corrections in any finite order. These are for example mass
differences among various heavy flavoured hadrons, inclusive width
differences between mesons and baryons or width splittings among the members
of multiplets, differences in spectra etc. These effects are
determined by
the differences in hadronic quantities like $\La$ for different hadrons;
after all they are expressed in terms of
the masses (or other characteristics)
of {\em hadrons} and are therefore unambiguous.
One note of caution should be added here:
in a quantum description it often happens that
inequalities lose their validity through a neccessary
subtraction procedure. For
example, the effective values of such quantities as $\bar Q
(i\ve{D})^2 Q$ -- though naively positively defined --
could in principle appear to
be of either sign.

In practice of course one deals with the real world of hadrons where the
masses of $c$ and $b$ quarks are rather large, but finite.
It is just in such a
situation when the analysis of the non-perturbative corrections
which scale like
$(1/m_Q)^n$ has practical relevance. It was shown that in many cases the
leading non-perturbative corrections are at
least as large numerically as the
perturbative
ones even for the beauty system \cite{BUV,BU2,BU3} and it is a practical
necessity to properly account for them. In particular the problem of
the
numerical value of $m_b$ and $m_c$, or in other words, $\La$ is important for
various phenomenological applications. On the other hand in practical
calculations one accounts only for a few terms in both perturbative and
nonperturbative expansions. This truncated procedure is justified for the
case of a finite mass $m_Q$: nonperturbative corrections are to account for
the physics of the low scale of order $\Lam$ whereas the perturbative
ones reflect the dynamics at high momenta $\sim m_Q$. The regime of
the perturbative corrections in a particular order may however cover in
reality also momenta that constitute
lower and lower fractions of $m_Q$; for a
fixed $m_Q$ at some order the two regions start to overlap thus violating
the physical basis for the two expansions. This suggests the conclusion that
for the real masses of $b$ and $c$ quarks non-perturbative
quantities including
$\La$ have a reasonable physical meaning, albeit with
an intrinsic limitation on
the possible accuracy of their values, both from a theoretical and
phenomenological point of view (see also \re{BSUV2} for a
similar discussion).
For example one may hope to get a reasonably
accurate {\em numerical} estimate of $\La$ from QCD sum rules (see e.g.
refs.~\cite{VLB,MN}) or lattice calculations at small, but finite
lattice spacing.
Needless to say that such a
quantity may depend significantly on the exact way
it is defined and to which order in the perturbative expansion
one has gone.

\vspace*{.2cm}
{\em To summarize}:
we have shown that the derivation of the bound on $\La$
obtained in ref.\cite{GM} is fundamentally flawed
if applied to the real physical quantity.
It actually may refer only to some different quantity rather than $\La$ for
which this inequality is satisfied trivially.
We have argued that in all likelihood
no rigorous bound of a similar type can be obtained in QCD. However
physically justified statements about $\La$ can in principle be obtained by
applying various dynamical consideration, with certain intrinsic limitations
on the possible numerical precision.

\vspace*{1cm}

{\bf ACKNOWLEDGEMENTS:} \hspace{.4em} N.U. gratefully acknowledges
illuminating discussions with A.Vainshtein during the
collaboration on many topics
that have relevance to the subject addressed in this paper, and to V.Yu.Petrov
for courageously sharing his doubts about a number of points where the
prevailing orthodoxy may seem to be preliminary.
This work
was supported in part by the National Science Foundation under grant number
PHY 92-13313.

\end{document}